\documentclass[%
aps,prd,nofootinbib,showkeys,a4paper,10pt
]{revtex4-2}

\usepackage{mathrsfs}          
\usepackage{graphicx}          
\usepackage{dcolumn}           
\usepackage{latexsym}
\usepackage{amsfonts}
\usepackage{amsmath}           
\usepackage{amssymb}
\usepackage{textcomp}          
\usepackage{bm}                
\usepackage{color}
\usepackage[utf8]{inputenc}    
\usepackage{hyperref}         

\begin{document}

\title{Variable Inflaton Equation of State and Reheating}

\author{Alessandro Di Marco}
\email{alessandro.dimarco.unifis@gmail.com}
\affiliation{ University of Rome - Tor Vergata, Via della Ricerca Scientifica 1}

\author{Gianfranco Pradisi}
\email{gianfranco.pradisi@roma2.infn.it}
\affiliation{ University of Rome - Tor Vergata, Via della Ricerca Scientifica 1\\
and INFN Sezione di Roma “Tor Vergata”, via della Ricerca Scientifica 1, 00133 Roma, Italy}


\begin{abstract}

We explore the consequences of a time-dependent inflaton Equation-of-State (EoS) parameter in the context of the post-inflationary perturbative Boltzmann reheating.
In particular, we numerically solve the perturbative coupled system of Boltzmann equations involving
the inflaton energy density, the radiation energy density and the related entropy density and temperature of the produced particle thermal bath.
We exploit reasonable Ans\"atze for the EoS and discuss the robustness of the Boltzmann system.
We also comment on the possible microscopic origin related to a 
time dependent inflaton potential, discussing the consequences on a 
preheating stage and the related (primordial) gravitational waves.

\end{abstract}



\maketitle

\tableofcontents

\section{Introduction}	

The slow-roll inflationary scenario \cite{1,2,3,4,5,6} is based on the introduction of a neutral, homogeneous and minimally coupled scalar field $\phi$,
the inflaton, usually equipped with an effective potental $V(\phi)$ characterized by an almost flat region and a fundamental vacuum state.
In the early phase of inflation 
the scalar field, displaced from the minima of its potential, 
slowly moves through the almost flat region of $V(\phi)$, covering a distance $\Delta\phi$ \cite{7} and mimicking a false vacuum or a ``quasi" de Sitter cosmology. 
As the scalar potential steepens, the inflaton falls in the global vacuum and starts to oscillate.
Being coupled to (extensions of) the Standard Model (SM) sector,  the inflaton is thus able to reheat the Universe
and to provide the graceful exit towards the initial standard radiation era of the Hot Big Bang cosmology 
(see \cite{8,9,10,11} for the first discussions about the reheating mechanism in slow-roll inflation). 
The properties of the reheating phase, however, strongly depend on the details of the slow-roll inflationary model, in particular on the form of the inflationary scalar potential. 
In the simplest examples, reheating occurs via the  
perturbative single body decay of the inflaton into the light SM (or Beyond the Standard Model (BSM)) particles, 
driven by the presence of interaction terms in the effective lagrangian. Typically, one has  familiar (renomalizable) three and four point interactions
or even higher order (non-renormalizable) interactions.  
Perturbative reheating allows the average energy density conversion to be modelled in terms of an effective Boltzmann perfect fluid.  
Despite its simplicity, the perturbative approach can be fruitfully used to characterize the evolution of the temperature of the relativistic plasma 
and the thermalization processes, together with the production of very heavy
particles, baryons, dark matter and/or dark radiation, 
even in cosmological models derived or inspired  by supersymmetry, supergravity or superstring theory \cite{12,13,14,15,16,17,18,19,20,21,22,23,24,25,26,27,28,29}.

Since the discovery of parametric resonance \cite{30,31,32,33,34,35}, 
it is known that reheating could be preceded by a non-perturbative preheating phase,
a rapid decay of the inflaton characterized by an exponential growing of the number of produced daughter particles.  
Preheating can be driven by different mechanisms, 
like the mentioned parametric resonance, tachyonic instability \cite{36} or a mixing of the two (tachyonic resonance \cite{37}),  
just to name a few. 
Important applications can be found in the context of plateau-like potentials \cite{38}.
Moreover, the preheating phase (if any) is generically followed by other non-linear phases, 
like a stochastic gravity wave emission \cite{39}, the oscillon formation \cite{40} and turbulent dynamics \cite{41}, 
before approaching the perturbative regime.  
Many review papers on the reheating mechanisms exist \cite{42}, while CMB-motivated reheating constraints can be found,  
{\it e.g.}, in refs. \cite{43,44,45,46,47,48}.

An important and common assumption generically used in the perfect fluid description is that the inflaton sector is 
naturally characterized by a constant Equation of State (EoS) parameter $w_{\phi}$.  
However, the fundamental physics at such early times is largely unknown, being the involved energy scale typically well above the TeV scale. 
A nontrivial $w_{\phi}$ could thus be more suited in order 
to capture the complexity of the effective dynamical evolution of the Universe, due to the coupling between the inflaton sector and the rest.
In this paper, we explore an inflaton sector characterized by a time dependent EoS, corresponding to (a sort of) an inflaton potential that varies during the reheating epoch.
The transition between different values of the EoS parameter
can occur earlier and faster during the non-perturbative era or, conversely, later and slower during the perturbative regime.
It can be mimicked in the perfect fluid description by guessing appropriate functional forms of $w_{\phi}$.  
We provide numerical solutions of the perturbative Boltzmann dynamics using two different Ans\"atze for the EoS. 
In particular, we show the evolution of the main macroscopic reheating variables, 
namely the inflaton energy density, the radiation energy density, the entropy density and, primarily, the temperature of the hot thermal bath.

The paper is organized as follows.  
In Sec. II we review the general picture of the standard perturbative Boltzmann-Einstein-Friedmann (BEF) description of reheating. 
In Sec. III we introduce some examples of effective time-dependent EoS and discuss their basic properties. 
We numerically solve the corresponding BEF equations for the relevant reheating variables, 
analyzing the dependence of the temperature behaviour on the functional form 
and on the corresponding free parameters entering the EoS functions.
In Sec. IV, we introduce a non-standard inflaton potential, 
discussing the qualitative consequences on the stochastic GW background from preheating.
Finally, Sec. V contains a summary and some perspectives.

In this document, we use the particle natural units $\hbar=c=1$, 
we indicate with $M_p=1/\sqrt{8\pi G_{N}}$ the reduced Planck mass,
where $G_{N}$ is the gravitational Newton constant, and we adopt the ``mostly minus" lorentzian metric signature $(+,-,-,-)$.

\section{Perturbative regime} 

According to the inflation paradigm, the early Universe has experienced  an effective cosmological phase characterized 
by a period of exponential expansion driven by one (fundamental, slow-rolling) weakly self-interacting scalar field, the inflaton $\phi$. 
The inflationary phase dilutes all the (pre-inflationary) relativistic entropies, energy and numerical densities, 
providing a very cold Universe.  
However, as the inflaton field rolls down and reaches the region where its effective potential $V(\phi)$ steepens, 
it falls and starts to coherently oscillate around the true vacuum,  
paving the way to the reheating process, whose properties are strongly model dependent.
In the single-field case, the inflaton is coupled to the SM field content or to its phenomenologically motivated BSM extensions. 
Consequently, one introduces a (cosmological) low energy effective action of the generic form
\begin{eqnarray}
\mathcal{S}\left[g_{\mu\nu},\phi\right]&\sim& S_{EH} + S_m \\ \nonumber
&=&\int d^4x\sqrt{-g} \left(-\frac{M_p^2}{2}R + \mathcal{L}(\phi,\chi,\psi)\right).
\end{eqnarray}
The first contribution $S_{EH}$ is the Einstein-Hilbert gravity sector, 
where $R$ is the Ricci scalar curvature and $g$ is the determinant of the (inflationary-induced) flat Friedmann-Lemaitre-Robertson-Walker (FLRW)
metric tensor $g_{\mu\nu}$. The line element associated to $g_{\mu\nu}$ reads
\begin{equation}
ds^2=dt^2 - a^2(t)dl^2, 
\end{equation}
with $t$ the cosmic time and $dl^2$ the euclidean line element of the metric of the three-dimensional spatial hypersurfaces at constant time. 
Finally, $a(t)$ is the standard cosmic scale factor.
The second contribution $S_m$ is the action of the matter sector,  where 
$\mathcal{L}$ is the lagrangian density describing the particle content of the post-inflationary phase of the Universe.  
Generically, it can be postulated to be of the form
\begin{align}
\mathcal{L}(\phi,\chi,\psi)&=\frac{1}{2}\partial_{\mu}\phi\partial^{\mu}\phi-V(\phi) \nonumber\\
&+\mathcal{L}(\chi)+\mathcal{L}(\psi)+\mathcal{L}_{int}(\phi,\chi) +\mathcal{L}_{int}(\phi,\psi) 
\label{eq:lag}\end{align}
where $V(\phi)$ is the effective potential of the inflaton field with a (curvature) mass term $m_{\phi}$ around the vacuum, 
while $\mathcal{L}(\chi)$ and $\mathcal{L}(\psi)$ describe the bosonic and fermionic sectors of the BSM, respectively.
Interactions are encoded within the $\mathcal{L}_{int}$'s terms where, in particular,  
$\mathcal{L}_{int}(\phi,\chi)$ contains the interactions between the inflaton and the bosonic BSM field content $\chi$, while
$\mathcal{L}_{int}(\phi,\psi)$ the ones between the inflaton
and the fermionic BSM field content $\psi$.
Notice that we are excluding, for simplicity, three-field interaction terms $\mathcal{L}_{int}(\phi,\chi,\psi)$ and we are  
also assuming that the scalar potentials, and in particular the hierarchy of masses, are protected against large radiative loop corrections, 
as it is customary in these kind of analyses.
In this section, we study the case in which the post-de Sitter phase of the Universe is basically characterized by a perturbative reheating 
with a very fast or completely negligible  preheating.
In the perturbative regime, the inflaton field oscillations are basically small ($\phi\ll M_p$) 
and the post-inflationary interactions allow single particle decays of the inflaton to bosonic and fermionic BSM fields, freely from tachyonic instabilities.
Once the specific structure of the interactions is known, it is possible to compute the total decay rate of the inflaton $\Gamma_{\phi}$
(as the sum of the decay rates for all decay channels) as well as the value of the reheating temperature, of the order of $\sim \sqrt{\Gamma_{\phi} M_p}$.
It is also quite interesting to study the global evolution of the post-inflationary reheating Universe, i.e. 
to analyse the way in which the energy density stored in the inflaton condensate is drained to the BSM fields.
In this respect, it is useful to exploit an effective perfect fluid description given by
the well-known system of coupled first order Boltzmann ordinary differential equations,  
able to describe the out-of-equilibrium decay of a scalar species into light relativistic particles, given by
\begin{eqnarray}\label{eqn: general system}
\begin{cases}
\dot{\rho}_{\phi} + 3H\gamma_{\phi}\rho_{\phi}=-\gamma_{\phi}\Gamma_{\phi}\rho_{\phi} \\
\dot{\rho}_{r} + 4H\rho_{r}=\gamma_{\phi}\Gamma_{\phi}\rho_{\phi}  \\ 
\dot{s}+3Hs=\gamma_{\phi}\Gamma_{\phi}\rho_{\phi}/T \\
\dot{T}+HT = \gamma_{\phi}\Gamma_{\phi}\rho_{\phi}/4\sigma T^3
\end{cases} \ .
\end{eqnarray}
In particular, $\rho_{\phi}$ is the (average) inflaton energy density, $\rho_{r}$ is the produced radiation energy density, $s$ is the entropy density
and $T$ is the temperature of the thermal bath.
The system must be paired with the Einstein-Friedmann equation
\begin{equation}
H^2=\frac{1}{3M^2_p}\left[\rho_{\phi}+\rho_r\right] 
\end{equation}
describing the evolution of the Hubble rate $H=\dot{a}(t)/a(t)$.
The dots label derivatives with respect to the cosmic time $t$.
Finally, $\gamma_{\phi}=1+w_{\phi}$, where $w_{\phi}$ is the effective EoS parametrizing the average properties
of the inflaton perfect fluid. In the case of a static inflaton potential $V \sim \phi^n$, one has $w_{\phi}=(n-2)/(n+2)$.  
For instance, the $n=2$ case corresponds to a matter-like pressure-less
inflaton with $w_{\phi}=0$, while the $n=4$ case matches a radiation-like inflaton with $w_{\phi}=1/3$ \cite{11,28,29}.
It should be stressed that in the system we also include the equation of evolution for the entropy density
and the temperature of the produced hot plasma of relativistic particles, based on the thermodynamic laws
\begin{equation}
\rho_r= \sigma T^4 ,
\end{equation}
where $\sigma=\pi^2 g_E/30$ is the Stefan constant, and 
\begin{equation}
s=\eta T^3 ,
\end{equation}
where $\eta=2\pi^2 g_S/45$. The quantity $g_E$ represents the number of effective relativistic degrees of freedom contributing to the energy,
\begin{equation}
g_{E}(T)=\sum_{b} g_{b}\left( \frac{T_b}{T}\right)^4 + \frac{7}{8}\sum_{f} g_{f}\left(\frac{T_f}{T}\right)^4,
\end{equation} 
while $g_S$ is the analogous quantity related to the degrees of freedom contributing to the entropy
\begin{equation}
g_{S}(T)=\sum_{b} g_{b}\left( \frac{T_b}{T}\right)^3 + \frac{7}{8}\sum_{f} g_{f}\left(\frac{T_f}{T}\right)^3.
\end{equation}
In both cases, $b$ and $f$ label bosonic and fermionic contributions, respectively. 
For temperature scales well above the QCD phase transition scale ($M_{QCD}\sim 200$ MeV)
we can safely assume $g_E, g_S \gtrsim 100$.  
For this reason, the ratio $g_S/g_E$ that should appear in the third equation of the system \eqref{eqn: general system} has been settled to one.
The system in Eq. \eqref{eqn: general system} describes the dynamics for $t\geq t_{end}$, where $t_{end}$ indicates the cosmic time at the end of inflation. 
Thus, one has to decorate it with a set of initial conditions. 
A natural choice is 
$\rho_{\phi}(t_{end})\simeq \rho(\phi_{end})$, $\rho_{r}(t_{end})\sim 0$,  $s(t_{end})\sim 0$ and  $T(t_{end})\sim 0$
\footnote{This assumption is reasonable in many inflationary scenarios, 
since the de Sitter phase dramatically dilutes all the numerical, energy and entropy densities initially present in the pre-inflationary Universe.}.
To simplify matters, it is useful to introduce the dimensionless ``time" variable $x=t/t_{end}$ with $x_{end}=1$ together with normalized reheating functions
with respect to (a proper power of) the energy density at the end of inflation, i.e.
$\bar{\rho_i}=\rho_i/\rho_{end}$ ($i=\phi,r$), $\bar{s}=s/\rho_{end}^{3/4}$ while $\bar{T}=T/\rho_{end}^{1/4}$.
As a result, one gets

\begin{eqnarray}\label{eqn:generalsystem_2}
\begin{cases}
\bar{\rho}_{\phi}' + 3H\gamma_{\phi}\bar{\rho}_{\phi}=-\frac{2k}{3}\Gamma_{\phi}\bar{\rho}_{\phi} \\
\bar{\rho}_{r}' + 4H\bar{\rho}_{r}=\frac{2k}{3}\Gamma_{\phi}\bar{\rho}_{\phi}  \\ 
\bar{s}'+3H\bar{s}=\frac{2k}{3}\Gamma_{\phi}\bar{\rho}_{\phi}/\bar{T} \\
\bar{T}'+H\bar{T} = \frac{2k}{3}\Gamma_{\phi}\bar{\rho}_{\phi}/4\sigma \bar{T}^3
\end{cases} \ ,
\end{eqnarray}

with 
\begin{equation}
H^2=\left(\frac{2}{3\gamma_{\phi}}\right)^2\left[\bar{\rho}_{\phi}+\bar{\rho}_r\right] , 
\end{equation}
where the prime indicates the derivative with respect to $x$.
In the resulting simplified notation, the initial conditions for the reheating variables read 
$\bar{\rho}_{\phi}(1)=1$, $\bar{\rho}_r(1)=0$, $\bar{s}(1)=0$ and $\bar{T}(1)=0$. 
A hierarchy parameter $k=\Gamma_{\phi}/H_{end}$ has been also introduced \cite{29}, that controls the duration of the reheat process, 
with the Hubble rate at the end of inflation given by $H_{end}\sim 2/(3\gamma_{\phi} t_{end})$. 
If the decay width is orders of magnitude smaller than $H_{end}$, then $k\ll 1$ and a prolonged reheating phase takes place after inflation, 
with the energy density budget of the Universe dominated by the inflaton vacuum modes.
In this limit, the solution of the Friedmann equation gives rise to a cosmic factor $a(x)$ that scales as
\begin{equation}
a(x)\sim a_{end} \ x^{2/3\gamma_{\phi}} ,
\end{equation} 
while the reheating functions result
\begin{eqnarray}\label{eqn:generalsystem3}
\bar{\rho}_{\phi}(x)&=& \frac{1}{x^2}e^{-\frac{2}{3}k(x-1)},\\
\bar{\rho}_r(x)&=& \frac{2k}{3} \frac{1}{x^{\alpha+1}}I_{\alpha}(x),\\
\bar{s}(x)&=& \eta \left(\frac{2k}{3\sigma}\right)^{3/4} \frac{1}{x^{3(\alpha+1)/4}}I^{3/4}_{\alpha}(x),\\
\bar{T}(x)&=& \left(\frac{2k}{3\sigma}\right)^{1/4}\frac{1}{x^{(\alpha+1)/4}}I_{\alpha}^{1/4}(x).
\end{eqnarray}
The integral $I_{\alpha}(x)$ involved in the previous equations is 
\begin{equation}\label{eqn: Boltzmann integral}
I_{\alpha}(x)=\int_{1}^x du \ u^{\alpha-1}e^{-\frac{2}{3}k(u-1)} , 
\end{equation}
with an $\alpha$ parameter depending on the EoS of the inflaton as
\begin{equation}
\alpha=\frac{8-3\gamma_{\phi}}{3\gamma_{\phi}} .
\end{equation}
In order to give an asymptotic behaviour of the integral, we can introduce the variable 
\begin{eqnarray}
v=\frac{2k}{3}u ,
\end{eqnarray}
in such a way that the integral becomes
\begin{eqnarray}
I_{\alpha}(\tilde{x})=\left(\frac{3}{2k}\right)^{\alpha}e^{\frac{2}{3}k} \int_{\tilde{x}_{end}}^{\tilde{x}} dv \  v^{\alpha-1} \ e^{-v},
\end{eqnarray}
where, of course,
\begin{equation}
\tilde{x}=\frac{2k}{3} x,\quad \tilde{x}_{end}=\frac{2k}{3}x_{end}.
\end{equation}
The integral can be decomposed as the difference of two integral functions \cite{28}
\begin{eqnarray}
I_{\alpha}(\tilde{x})=\left(\frac{3}{2k}\right)^{\alpha}e^{\frac{2}{3}k}\left[\Omega_{\alpha}(\tilde{x}) - \Omega_{\alpha}(\tilde{x}_{end})\right],
\end{eqnarray}
where $\Omega_{\alpha}(\tilde{x})$ is the lower incomplete Euler Gamma function $\gamma(\alpha, \tilde{x})$ \cite{49}, 
whose power series expansion results
\begin{eqnarray}
\gamma(\alpha, \tilde{x})=\sum_{n=0}^{\infty} c_n \tilde{x}^{\alpha+n}  ,
\end{eqnarray}
with the expansion coefficients given by 
\begin{equation}
c_n=\frac{(-1)^n}{n!(\alpha+n)}.
\end{equation}
The integral can thus be written as 
\begin{eqnarray}
I_{\alpha}(\tilde{x})=\left(\frac{3}{2k}\right)^{\alpha}e^{\frac{2}{3}k}\sum_{n=0}^{\infty}c_{n}\tilde{x}^{\alpha+n}
\left[ 1-\left(\frac{\tilde{x}_{end}}{\tilde{x}}\right)^{\alpha + n} \right] .
\end{eqnarray}
In our case, since $k\ll1$, we can limit ourselves to a region  where $\tilde{x}_{end}\ll \tilde{x} \ll 1$. 
As a consequence, the dominant term is the one with $n=0$, corresponding to
\begin{eqnarray}
I_{\alpha}(\tilde{x})\sim \left(\frac{3}{2k}\right)^{\alpha}e^{\frac{2}{3}k} \frac{\tilde{x}^{\alpha}}{\alpha}\left[ 1-\left(\frac{\tilde{x}_{end}}{\tilde{x}}\right)^{\alpha} \right].
\end{eqnarray}
Remembering the definition of $\tilde{x}$ in terms of $x$, we obtain the first order expression of Eq.($\ref{eqn: Boltzmann integral}$)
\begin{equation}
I_{\alpha}(x)\sim \frac{x^{\alpha}}{\alpha}\left(1-\frac{1}{x^{\alpha}}\right) ,
\end{equation}
so that 
\begin{eqnarray}\label{eqn:generalsystem4}
\bar{\rho}_{\phi}(x)&=& \frac{1}{x^2}e^{-\frac{2}{3}k(x-1)},\\
\bar{\rho}_r(x)&=& \frac{2k}{3\alpha}\frac{1}{x}\left(1-\frac{1}{x^{\alpha}}\right),\\
\bar{s}(x)&=&  \frac{\bar{s_0}}{x^{3/4}}\left(1-\frac{1}{x^{\alpha}}\right)^{3/4},\\
\bar{T}(x)&=&  \frac{\bar{T_0}}{x^{1/4}}\left(1-\frac{1}{x^{\alpha}}\right)^{1/4},
\end{eqnarray}
with coefficients $\bar{s}_0$ and $\bar{T}_0$ defined as
\begin{eqnarray}
\bar{s}_0= \frac{2\pi^2 g_E}{45} \left(\frac{20k}{\alpha\pi^2g_E}\right)^{3/4},   
\quad \bar{T}_0= \left(\frac{20k}{\alpha\pi^2g_E}\right)^{1/4}.
\end{eqnarray}
The reheating phase is characterized by peaks in the temperature and entropy sectors that occur at quite early times. The process 
tends to become complete as the Hubble rate reaches the $\Gamma_{\phi}$ scale, 
corresponding to a normalized time epoch $x\sim x_{reh}\sim k^{-1}$ \cite{29}.
In particular, the inflaton energy density decreases exponentially close to $x_{reh}$ producing most of the relativistic particles 
and the Hot Big Bang cosmology begins with an initial temperature, the reheating temperature, of the order of $T_{reh}= c \sqrt{\Gamma_{\phi}M_p}$, 
where the prefactor $c$ depends on the details of the inflaton oscillation \cite{29}.

\section{Perturbative regime and nonstandard Equation of State}

The predictions of the previous standard physical scenario are good and well-motivated and 
could be finally improved by adding a (possible) preliminary preheating stage.
Nevertheless, the physics at such high energy scales remains substantially unknown 
and non-trivial mechanisms could enter the game with interesting consequences.  
Indeed, the perturbative regime should be also supplemented by modifications due to the complexity of the reheating phase. 
First of all the interaction between the inflaton, the BSM fields and even other fundamental scalar fields (moduli) could be governed by couplings
giving rise to a global time-dependent inflaton potential about the post-inflationary vacua (we will show an example in the last section).
Moreover, one should also take into account other possible non-linear effects.
For instance, it is obvious that backreaction and re-scattering within the BSM matter production process should play an important role.
In order to capture all these possible features, it is conceivable to generalize the  ``effective" EoS parameter $w_{\phi}$ of the inflaton field to be time-dependent.
A proposal could be to model $w_{\phi}(t)$ as evolving, for example, from $w_{\phi}=0$ to $w_{\phi}=1/3$.  
In this way, we expect to get a behavior of the reheating functions different from the one emerging by the analytical solutions mentioned before.  
First of all, an analytic solution of the system with a time-dependent $w_{\phi}$, generically, is no longer at disposal. 
Second, the evolution of the system will depend upon the Ans\"atze for the scalar function $w_{\phi}$, 
i.e. on the way in which the inflaton oscillations change in time.
For these reasons, a pure numerical approach appears to be the best choice to address the problem.  
Of course, a good prescription for the time evolution of the inflaton EoS must be introduced.
Several possibilities are allowed in order to model in a reasonable way an evolution from $0$ to $1/3$.  
In this paper, we explore the impact of two possible dynamics: a smooth evolution and an oscillatory evolution.

\subsection{Rayleigh-Weibull Equation-of-State}

The simplest way to connect two different regimes of the EoS is via a smooth evolution or a ``cumulative" function.
For example, one could implement $w_{\phi}$ with an arctangent form
\begin{equation}
w_{\phi}(x)=\frac{1}{6}\left[ 1 + \frac{2}{\pi}\arctan(x) \right]
\end{equation}
or with an integral form by using the Error function
\begin{equation}
w_{\phi}(x)=\frac{1}{6}\left[ 1 + \mbox{Erf}(x)\right] .
\end{equation}
However, the problem with these Ans\"atze is that they are not enough flexible, since they strongly constrain the evolution from the initial value to the final one.
Therefore, a possible and rather different possibility is to use a scalar function equipped with adjustable parameters.
Our choice is to introduce a generalization of the Rayleigh-Weibull (cumulative) function \cite{50}
\begin{equation}\label{eqn: w}
w_{\phi}(x)=\frac{1}{3}y(x), \quad y(x)= 1 - e^{-f(x)} . 
\end{equation}
with
\begin{equation}
f(x)=\left(\frac{x-\mu}{\sigma}\right)^p.
\end{equation}
The presence of the three free parameters $p$ (positive), $\mu$ and $\sigma$,
allows us to model several behaviours of $w_{\phi}$ (for $x\geq \mu$) from $0$ to $1/3$ within the same dynamical framework.
Indeed, for small time scales $x\sim \mu$ one gets $f(x)\sim 0$, so that $w_{\phi}\sim 0$ as well.
On the other side, the limit of very large time scales $f(x)\sim (x/\sigma)^p \gg 1$ (with $x<x_{reh}$, {\it i.e.} $\sigma \ll x <  x_{reh}$)  
provides $w_{\phi}\sim 1/3$ before the reheating completion.
However, nothing prevents to relax the last condition in order to get a value of the EoS at $x_{reh}$ smaller than $1/3$.
In Fig.(1) the evolution of the EoS for some $p$ values and $\sigma=5, \mu=1$, is reported.  
Notice that all the curves meet at the point $x_0= \mu + \sigma$, being $f(x_0)=1$ independently of $p$. The parameter $p$ strongly determines the shape of $w_{\phi}$.
Indeed, for $p<1$ the scalar function assumes a monotonic behaviour 
from the origin of the dynamics, ($x=\mu$), and $w_{\phi}$ grows slowing without inflection points.  
On the contrary, for $p>1$ the EoS tends to be much more steep, reaching the more rapidly the radiation regime the higher the value of $p$ is. 
In particular, two inflection points are present and the latest at
\begin{equation}\label{eqn: inflection point}
x_{eff}=\mu +\sigma\left(\frac{p-1}{p}\right)^{1/p}
\end{equation} 
characterizes the effective critical epoch at which the effective EoS parameter inverts the kind of growth towards the end of the reheating phase.  
Finally, $\sigma$ is inserted to tune the time scale with respect to $x_{reh}$.

\begin{figure}
\centering
\includegraphics[width=12cm, height=8cm]{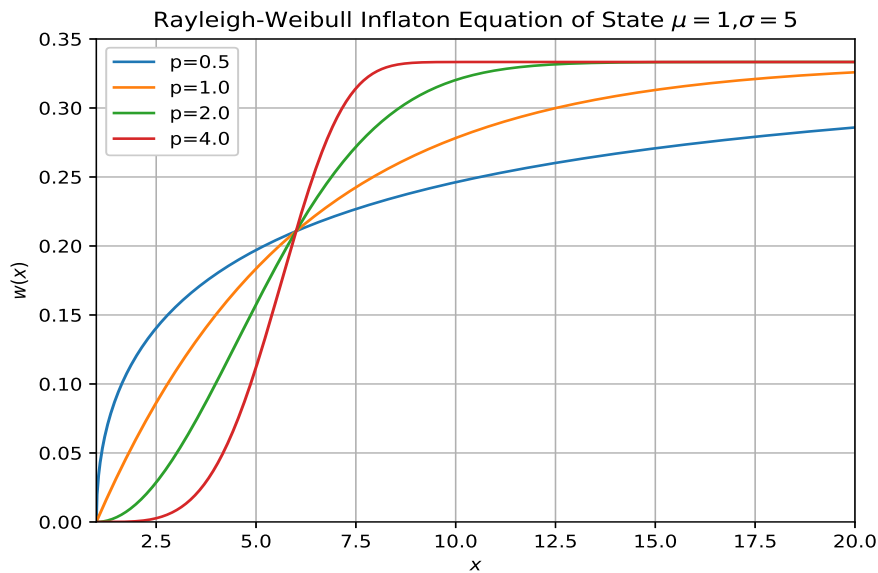}
\caption{Dynamics of the inflaton Equation-of-State $w_{\phi}$. In the limit of small $p$ the $w_{\phi}$ is a monotonically increasing function
with no inflecion points while for $p$ larger then unity, it develops an inflection point as given by Eq.($\ref{eqn: inflection point}$) that
represents the maxima of the $w'(x)$.}
\label{fig: 1}
\end{figure}

In the presence of a time-dependent EoS the coupled system can be written in the familiar form 
\begin{eqnarray}\label{eqn:generalsystem_cont}
\begin{cases}
\bar{\rho}_{\phi}' + 3H\gamma_{\phi}\bar{\rho}_{\phi}=-\frac{2k}{3}\Gamma_{\phi}\bar{\rho}_{\phi} \\
\bar{\rho}_{r}' + 4H\bar{\rho}_{r}=\frac{2k}{3}\Gamma_{\phi}\bar{\rho}_{\phi}  \\ 
\bar{s}'+3H\bar{s}=\frac{2k}{3}\Gamma_{\phi}\bar{\rho}_{\phi}/\bar{T} \\
\bar{T}'+H\bar{T} = \frac{2k}{3}\Gamma_{\phi}\bar{\rho}_{\phi}/4\sigma \bar{T}^3
\end{cases} \ ,
\end{eqnarray}
where now
\begin{equation}
\gamma_{\phi}(x)=1+\frac{1}{3}y(x)
\end{equation}
and
\begin{equation}
H^2(x)=\left(\frac{2}{3}\right)^2\left[\bar{\rho}_{\phi}(x)+\bar{\rho}_r(x)\right].
\end{equation}
It is important to notice that $\Gamma_{\phi}$ now plays the role of an effective (constant) decay rate. 
Moreover, the cosmic time $t$ is normalized with respect to $t_{end}\sim 2/3H_{end}$ because of the  initial matter-like inflaton, with $w_{\phi}=0$. 
In Fig.(2) we summarize the numerical solution \cite{52} of the Cauchy problem for the four fundamental reheating functions, 
by setting a reheating completion at time scales of the order of $x_{reh}=k^{-1}\sim 10^2$ and $p=10$.
As shown, the average energy density of the inflaton sector tends to decrease quite more rapidly than in the standard scenarios, 
starting at a certain epoch that corresponds to $x\sim 30-40$ in our numerical example. Around the reheating epoch, 
the energy density is significantly smaller than the cases with $w_{\phi}=0$ or $w_{\phi}=1/3$. As a consequence, 
the energy density of the produced thermal bath exhibits a little bump, immediately converted 
in the evolution of the remaining observables.

\begin{figure}
\centering
\hspace*{-1cm}
\includegraphics[width=14cm, height=12cm]{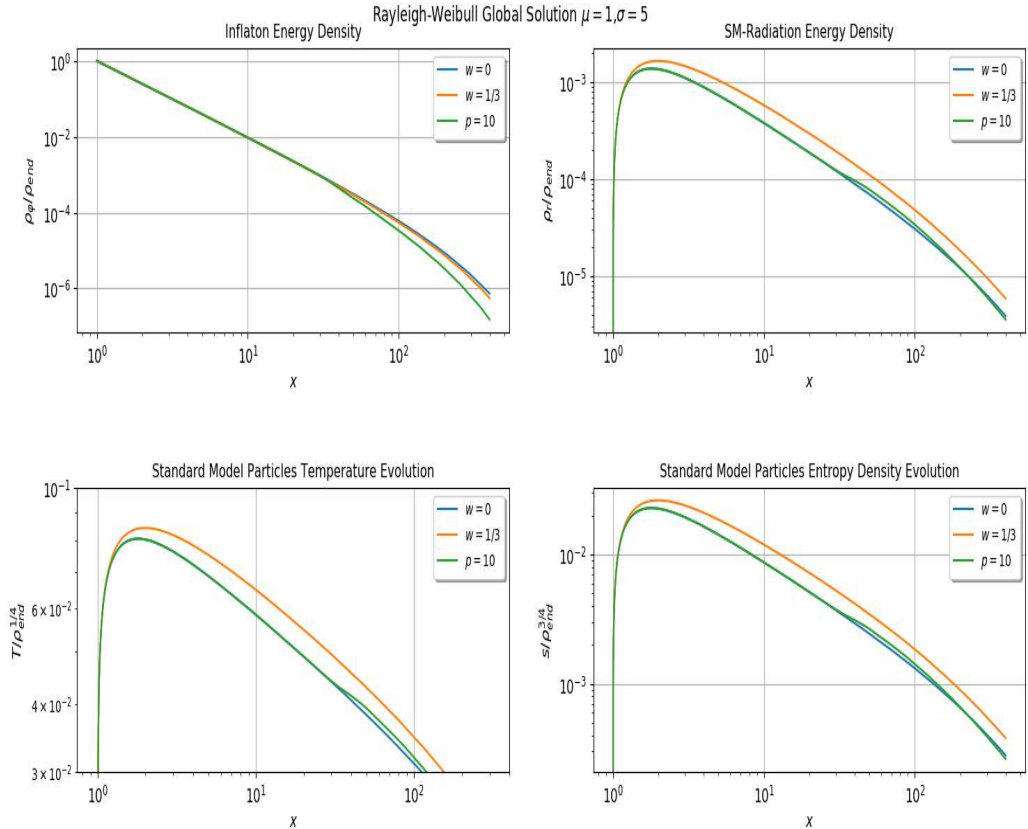}
\caption{Solution of the coupled system of Boltzmann equations. 
The reheating epoch is fixed to $x_{reh}\sim 10^2$ while the free EoS parameter is $p=10$.}
\label{fig: 2}
\end{figure}

In Fig.(3) we emphasize the evolution of the temperature of the relativistic plasma of BSM particles.
The EoS for small values of the parameter $p$ quickly moves away from the matter scenario ($w_{\phi}=0$), slowly growing towards the radiative case when time increases. 
This behaviour determines the temperature evolution at small times.  Indeed, the related temperature curves (green and red) tend to depart from the (blue) curve, 
that represents the temperature background
in the case of a matter-like inflaton, trying to follow the (orange) curve describing a pure radiation-like inflaton.
As a result, the maximum temperature reached in these kind of scenarios is a little higher than in the region with $p\geq 1$.
In the $p>1$ case, on the contrary, the EoS grows following a logistic-like curve, that approaches a step (or Heaviside) function in the limit of $p\gg 1$. 
As a consequence, the function $w_{\phi}$ remains arbitrarily close to that of a matter-like scenario quite for a long time. 
Therefore, the corresponding temperature curves (brown and pink) substantially match the matter case as well, 
displaying almost the same maximum temperature and deviating from the matter-case curve only at large time scales.
In the Rayleigh scenario ($p\sim 1$), the departure from the $w_{\phi}=0$ case begins at $x\sim\sigma$ and proceeds in a smooth way.  
For the $p\sim 10$ or the $p\sim 100$ cases, the transition naturally occurs at larger times and gets more and more striking (with a clear bump) as $p$ gets larger and larger.
Let us just mention that it is also possible to consider an ``inverted behaviour'' for the evolution of the EoS parameter, like 
\begin{equation}\label{eqn: inverted}
w(x)\sim\frac{2/3}{1+e^{f(x)}}  ,
\end{equation}
where the function $f(x)$ is the same as before.
In this case, one has a specular dynamics with the EoS emerging from inflation at a value close to $1/3$
and asymptotically approaching the pressureless matter case.  Moreover, we could consider an initial condition somewhat different from $1/3$, 
for example at a value $1/2$, related to an inflaton potential $\phi^6$. These kind of scenarios naturally imply an higher maximum reheating temperature, 
providing a richer high energy particle physics phenomenology, with the subsequent dynamics governed by the standard case with a $w_{\phi}=0$.  
Needless to say that it would also be possible to consider
situations in which the asymptotics after reheating is not given by  pure  matter or pure radiation fluids, but involves some intermediate exotic cases.

\begin{figure}
\centering
\hspace*{-1cm}
\includegraphics[width=15cm, height=10cm]{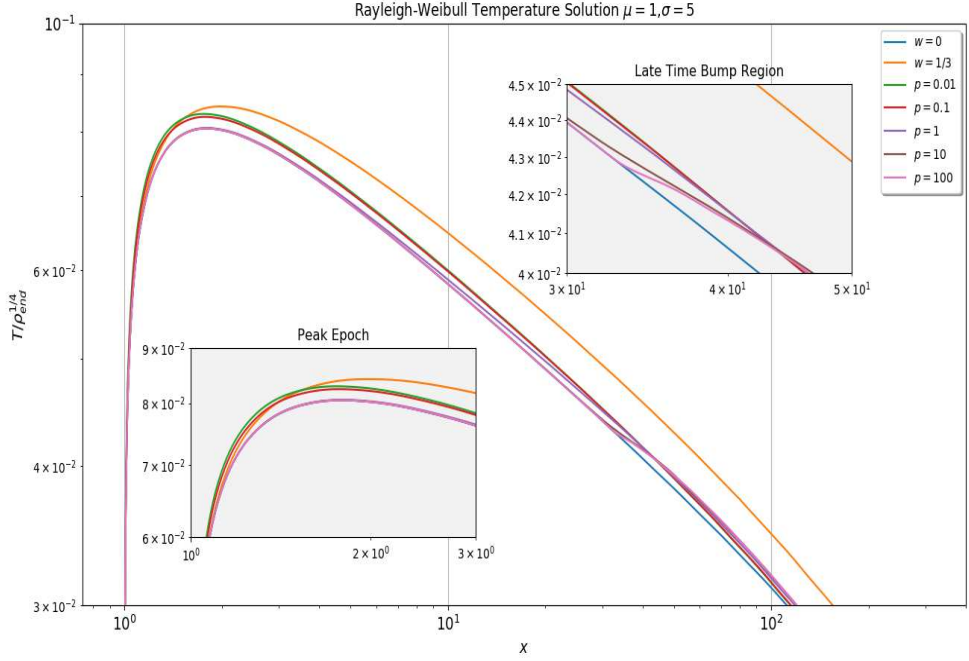}
\caption{Temperature evolution of the BSM relativistic plasma for different values of the parameter $p$}
\label{fig: 3}
\end{figure}

\subsection{Oscillatory Equation-of-State}

In this section, we introduce an EoS with oscillatory evolution in order to simulate and/or catch up some exotic 
(although regular) behaviour in the EoS inflaton cosmological sector.   
In particular, we adopt the same form of the scalar function
\begin{equation}
w(x)=\frac{1}{3}y(x), \quad y(x)= 1 - e^{-f(x)} ,
\end{equation}
selecting in this case
\begin{equation}
f(x)=\left(\frac{x-A\sin(px)}{\sigma}\right)^2 .
\end{equation}
The parameter $A$ is the amplitude of the oscillation, $p$ is the frequency of the oscillation while the parameter $\sigma$
tunes the dilution of the dynamics with time.  
In order to still obtain a matter-like inflaton around the end of inflation $x_{end}\sim 1$, one has to require $f(x)\ll 1$ or
\begin{equation}\label{eqn: small time prescription}
\delta\sim\left(\frac{1 - A\sin(p) }{\sigma}\right)^2\ll 1.
\end{equation}
The trigonometric function is bounded in the interval $[-1,1]$. 
However, one could have large oscillation with $A\sin(p)\gg 1$.  
In this case, the condition can be turned into
\begin{equation}
\delta\sim\frac{A^2\sin^2(p)}{\sigma^2}\ll 1.
\end{equation}
Moreover, the limit of large time scales must give rise to $w_{\phi}\sim 1/3$ or $f(x)\gg 1$, that amounts to
\begin{equation}\label{eqn: large time prescription}
\left(\frac{x}{\sigma}\right)^2\gg 1.
\end{equation}
We expect that these kind of EoS scenarios provide a fluctuation pattern on the evolution of the temperature background, 
characterized by two important properties: the magnitude of the fluctuations and the time scale where the fluctuations weaken.
For instance, if we set a reheating completion again at $x_{reh}\sim 10^2$, 
we get fluctuations around the peak for $\sigma^2\sim 1-15$ and fluctuations at large time scales for $\sigma^2\sim 10^2-10^3$.
In addition, as $x^2/\sigma^2\sim 1$ the fluctuation pattern starts to be suppressed.
In Fig.(4) we report an example of the oscillatory EoS behaviour that can provide a fluctuation pattern around the peak.
In particular, we use three choices of the parameter $\sigma^2$ with $p=8$ and $A=1$.
On the other side, in Fig.(5) we report a prototype example useful to induce fluctuations at later times.
In this case we set larger values of $\sigma^2$ with $p=8$ and $A=2$.
In order to study the cosmological evolution, we focus on small time-scale oscillations around the peak (with $p=8$), 
considering the different pairs $(A=1,\sigma^2=5)$, $(A=2,\sigma^2=10)$, $(A=3,\sigma^2=15)$. 
It should be noticed that the logistic type of variation is more and more evident as $\sigma$ becomes larger and larger. 

The corresponding numerical integration of the Boltzmann system provides the temperature evolutions shown in Fig.(6).  
A couple of interesting observations emerges.  
The number and the amplitude of fluctuations increases with the increasing of the
parameter $A$, given a fixed value of the ratio $A/\sigma^2$. 
Moreover, the extension of the time interval after which there is a damping of the fluctuations is proportional to the values of $\sigma^2$.  
In the limit of high energy scales, these fluctuations in the thermal bath can have a non-trivial impact. 
For instance, they can modify the production of heavy relic particles and their final abundances.

Our examples are constrained by the initial prescription, i.e. the choice of having $w_{\phi}\sim 0$
at the beginning of the reheating phase and $w_{\phi}\sim 1/3$ at the end of it.  
Clearly, there is room for a plethora of other possibilities. 
They can involve different initial or final conditions on $w_{\phi}$, 
a time-dependent behaviour that allows a plateau-like temperature peak able to guarantee a more efficient particle production, 
a total decay amplitude $\Gamma_{\phi}$ also variable in time, and so on.
Apparently, however, the perturbative Boltzmann system appears to be robust with respect to perturbations of the EoS, 
in the sense that the fluid tends to absorb the fluctuations allowing an asymptotic convergence of the evolution towards the radiation dominated epoch,
provided the amplitude of fluctuations does not destroy the perfect fluid approximation.

\begin{figure}
\centering
\includegraphics[width=12cm, height=8cm]{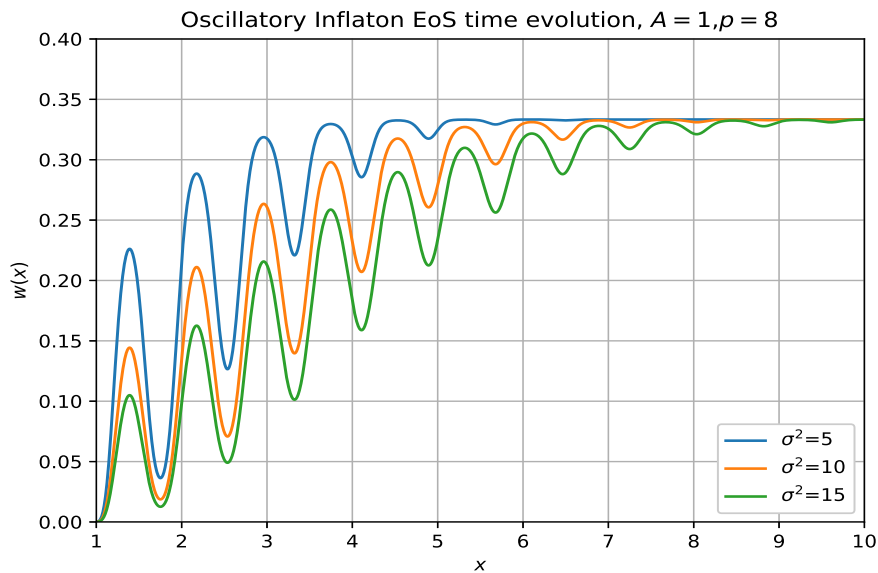}
\caption{Oscillatory Equation of State I. }
\label{fig: 4}
\end{figure}

\begin{figure}
\centering
\includegraphics[width=12cm, height=8cm]{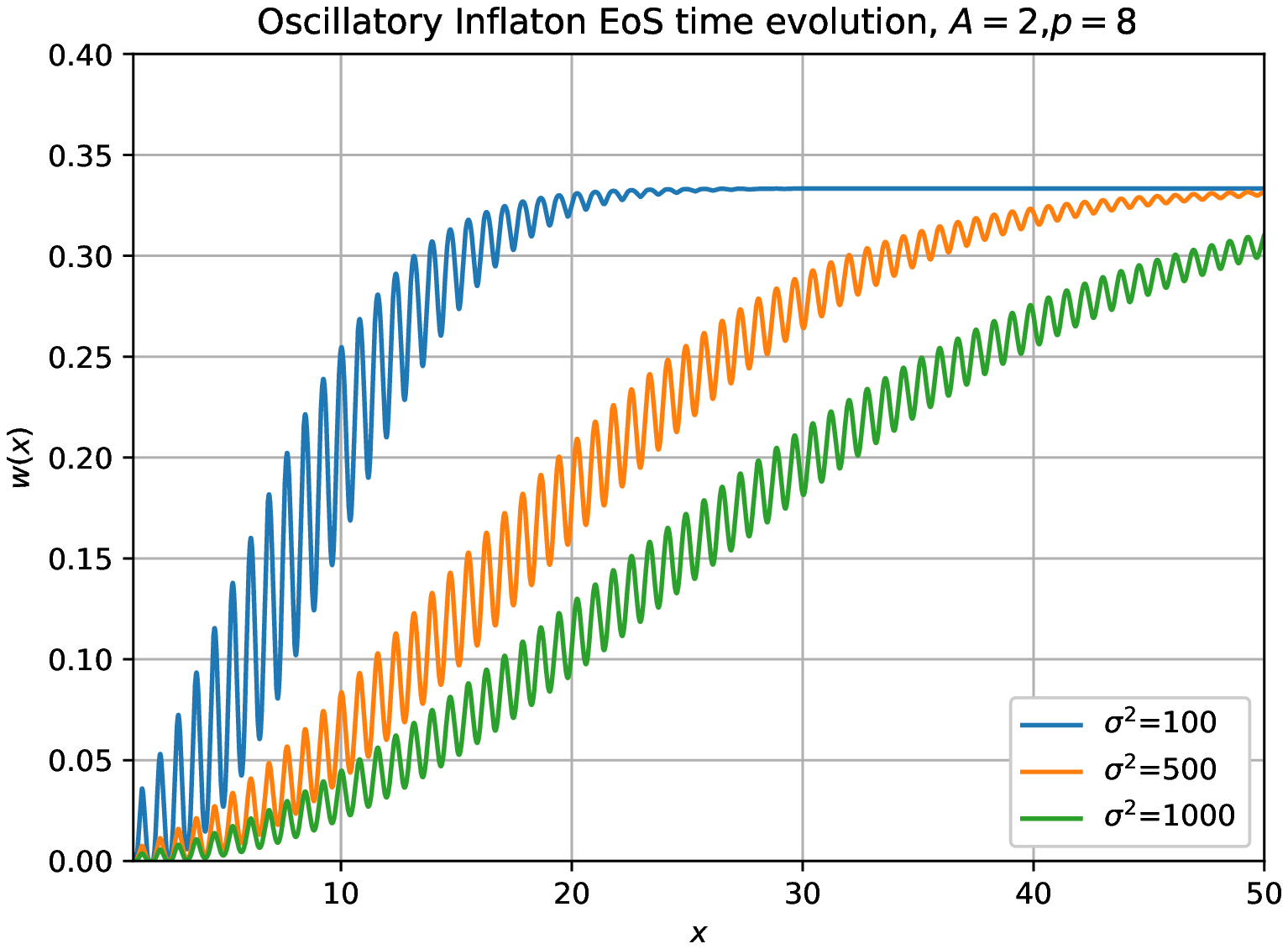}
\caption{Oscillatory Equation of State II.}
\label{fig: 5}
\end{figure}

\begin{figure}
\centering
\hspace*{-1cm}
\includegraphics[width=15cm, height=10cm]{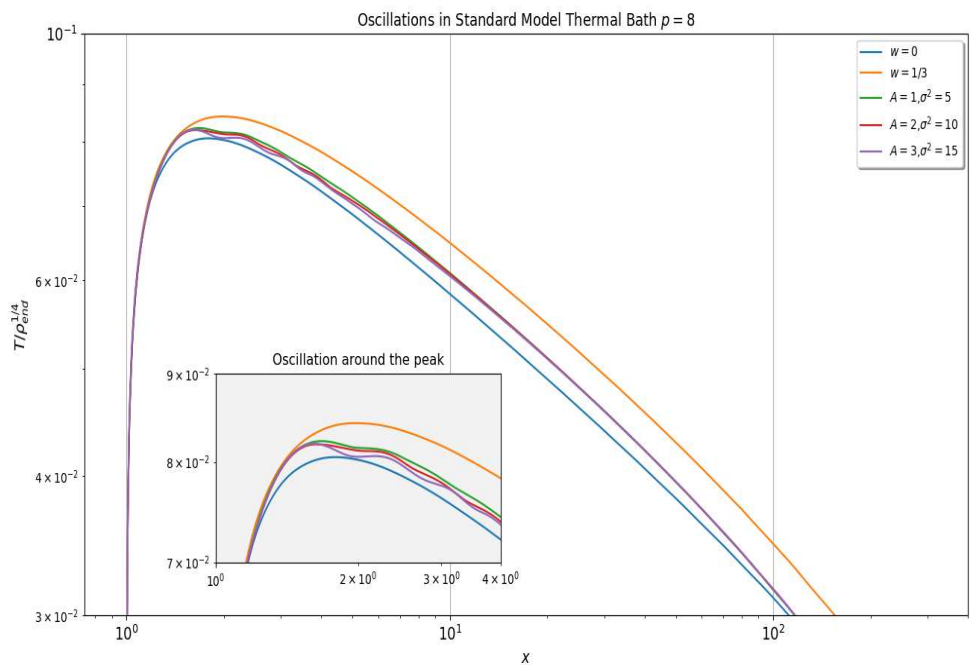}
\caption{Temperature evolution of the BSM relativistic plasma for different oscillatory behaviour of the inflaton EoS.}
\label{fig: 6}
\end{figure}

\section{Non Perturbative Regime}

In the previous sections, we have discussed a standard perturbative (perfect fluid) reheating cosmology,
by analyzing the effects of the inclusion of a suitable time dependence in the EoS parameter $w_{\phi}$.
However, it is well known that the reheating epoch can be characterized by a preliminary preheating stage, 
{\it i.e.} a rapid and non-perturbative explosive particle production mechanism, 
followed by the standard perturbative regime and the related thermalization era.
The explosive particle creation is due to an amplification of the mode-functions of the field(s) directly coupled to the inflaton.
The amplification mechanism can be realized in many ways
and the specific details depend on 
the underlying elementary particle theory and on the inflationary model.  
Both the shape of the inflationary scalar potential and 
the structure and magnitude of the inflaton couplings to the light fields of the visible (and/or hidden) matter sector(s) influence the preheating stage.  
In particular, possibilities are the production of (primordial) gravitational waves of classical quadrupole nature
or generated by the collision of long-lived pseudo-topological structures (oscillons).  
In addition, a potential of the form  
\begin{equation}
V(\phi)\sim \frac{1}{2}m^2_{\phi}f_1(\theta_i)\phi^2 + \frac{\lambda_{\phi}}{4}f_2(\theta_i)\phi^4 
\label{eq:timepotential}\end{equation}
is conceivable, where the dimensionless $f_1$ and $f_2$ allow a mass term domination at early times and a quartic self-interaction term domination at late times.  
Effective potentials of this sort are quite natural in string theory, where the effective couplings depend on the vev's of (stabilized or running) moduli, 
generically indicated with $\theta_i$ in Eq. \eqref{eq:timepotential}. 
A similar effective structure could fully justify an averaged EoS parameter $w_{\phi}$ varying with time.  
The simple interpretation in terms of the geometry of the inflaton vacuum is that 
in the limit of small cosmic times, the geometry of the vacuum is basically quadratic while it smoothly turns to a quartic one as time goes by.

The dynamics of the non-perturbative amplification mechanisms of pivot $k$-modes of (minimally coupled) light fields $\chi$ during preheating depends, 
of course, on the interaction with the inflaton.
In the case of a simple four-legs 
interaction of the form $g^2\phi^2\chi^2$
(parametric resonant preheating), for instance, 
it is well known that the equation regulating the momentum modes of the $\chi$ fields is generically dependent on the inflaton oscillation. 
The growing and evolution of modes result in a very complicated superposition of different resonant components, 
together with the expansion of the universe and a certain amount of lack of homogeneity of the inflaton field itself.  
The amplification of modes leads to inhomogeneous and time-dependent energy density and, unavoidably, to the generation of gravitational waves.  
Their spectra have been inferred in many examples \cite{39} and of course a non-trivial potential like the one 
in Eq. \eqref{eq:timepotential} significantly affects them, 
with properties and details related to the form and the behaviour of the functions $f_1$ and $f_2$.  
It would be very interesting to understand quantitatively, even in simple cases, the deviation from the standard predictions.  
The whole picture still holds even in more peculiar cases like quintessence scenario, where an additional (fundamental?)
scalar field equipped with a similar potential can be responsible for the reheat of the Universe.
Moreover, the non-trivial shape of the inflaton potential inevitably affects the formation of oscillons and their collisions, 
usually responsible for the production of an additional background of GW.
We defer a deep analysis of these aspects to future work.

\section{Summary and Conclusions}

In this paper we explored a cosmological scenario characterized by an inflaton time-dependent EoS during
the perturbative Boltzmann reheating stage in the regime from $w_{\phi}=0$ to $w_{\phi}=1/3$.
In particular, we modelled the time evolution of the EoS via two particular Ans\"atze: a generalization of the cumulative Rayleigh-Weibull function and an oscillating function.
In both cases we studied some cosmological consequences, with particular emphasis on the time evolution
of the temperature backgrounds relative to the hot gas of particles for given values of the varying EoS parameter.
The numerical integration of the Boltzmann system demonstrates the formation of bumps and patterns of oscillations 
in the temperature evolution that significantly affect the reheating phase, modifying the related cosmological parameters.
The time variation of the effective inflaton EoS can be ascribed to a non-trivial inflationary potential, dominated around the minimum 
by different potential terms in different epochs, that modify the coherent oscillations of the inflaton.  
In the case of a preheating phase, 
the non-standard inflaton potential also influence the unavoidable production of (primordial) gravitational waves, 
giving rise to peculiar spectra from the superposition of momentum modes of the produced matter fields. 
Moreover, the same source can affect generation and collision of oscillons together with the related GW background.
We aim at better investigating the consequences of the additional inhomogeneities in forthcoming papers.

\section*{Acknowledgments}

We would like to thank P.Cabella for the valuable comments on the early version of the manuscript and
Giancarlo De Gasperis for useful discussions about numerical integration aspects touched upon in this paper.

\bibliography{apssamp}

\end{document}